\documentclass[11pt,a4paper]{article}

\usepackage{amsmath,amssymb,graphicx,cite}
\usepackage{hyperref}
\hypersetup{
    colorlinks=true,
    linkcolor=blue,     
    urlcolor=cyan,
    }

\usepackage{subcaption}

\begin{document}

\title{Shadows cast by a class of rotating black bounces with an anisotropic fluid} 
\author{Ernesto F. Eiroa\thanks{e-mail: eiroa@iafe.uba.ar}, Juan M. Paez\thanks{e-mail: jmpaez@iafe.uba.ar}\\
{\small  Instituto de Astronom\'{\i}a y F\'{\i}sica del Espacio (IAFE, CONICET-UBA),}\\
{\small Ciudad Universitaria, 1428, Buenos Aires, Argentina}} 
\date{}
\maketitle

\begin{abstract}

In this work, we introduce a family of rotating black bounces with an anisotropic fluid obtained by using a modified Newman-Janis algorithm. We analyze the main features of these spacetimes and obtain the geodesics for photons, which admit the separation of the Hamilton-Jacobi equation. We then determine the shape of the shadow in terms of the parameters of the model. We present some relevant examples that exhibit some traits that distinguish these spacetimes from other related ones appearing in the literature.

\end{abstract}

\section{Introduction}\label{sec:intro}

The theory of general relativity predicts that the trajectories of light rays coming from the accretion disk or other sources behind a black hole are gravitationally deflected, resulting in a region in a distant observer sky from which no light arrives, usually denominated shadow \cite{bardeen1972,luminet1979,falcke2000}. In the last few years, the Event Horizon Telescope (EHT) Collaboration has obtained the reconstructed images of supermassive black holes M87* \cite{eht2019,eht2024}, at the center of the elliptical galaxy M87, and of Sgr A* \cite{eht2022}, at the center of our Galaxy. In these images, a dark region is surrounded by a bright ring of light, with an angular diameter of about $50\, \mu \mathrm{as}$. In general, the size and shape of the shadow depend on the distance and the inclination angle of the observer, and also on the parameters characterizing the black hole. For the Kerr spacetime of Einstein gravity, these quantities are the mass and the angular momentum of the black hole. When general relativity is coupled to additional fields, the size and shape of the shadow differ from the Kerr case, having a dependence on the extra parameters of the model considered. In modified theories of gravity, the features of black hole shadows also change. The study of the images of black holes then provides a way of distinguishing Einstein gravity from other possible scenarios. Analytical studies on black hole shadows rely on the separability of the Hamilton-Jacobi equation for photons; see \cite{cunha2018,perlick2022} for reviews. Among the large amount of interesting works in the literature, one can mention those concerning shadows in Einstein gravity, in vacuum \cite{hioki2009,tsukamoto2014,tsupko2017,li2020} or including fields or fluids \cite{hioki2008,amarilla2013,hou2018,badia2020,other2020,konoplya2021,heydari2022}, and also in theories of modified gravity \cite{tsukamoto2018,shaikh2019,cusped_3,kumar2020,bw1,bw2,bw4,other2010,other2021a,other2021d,other2022a,other2022b,sarikulov2022,sun2022, sengo2023,liu2024, galtsov2024,li2024,he2025}. Astrophysical black holes are expected to be surrounded by a plasma medium; then several studies that analyze how the presence of the plasma affects the characteristics of the shadow have appeared in recent years \cite{perlick2015,perlick2017,yan2019,babar2020,lima2020,fathi2021,zhang2023,badia2021,bezdekova2022, briozzo2022,badia2023,kobialko2024,perlick2024}. In this case, the properties of the shadow are chromatic, because the effect of the plasma on the light propagation depends on the frequency. In particular, for low enough frequencies, the black hole shows a ``forbidden region'' in which light cannot enter, resulting in an important decrease of the shadow size \cite{perlick2017,perlick2024}. The Hamilton-Jacobi equation for light rays is still separable when the density of a pressureless and nonmagnetized plasma obeys a suitable condition \cite{lima2020}. Shadows produced by naked singularities show distinct features \cite{joshi_2020,kerr_naked}. However, for systems that are not completely integrable, the shadows can only be obtained numerically \cite{Wang:2022kvg}; in this regard, one can point out the articles \cite{other2015,pyhole,cusped_4,Wang:2018eui,Wang:2021ara,Bronzwaer:2021lzo,Garnier:2023lph,Chen:2023wzv,Moreira:2025ckc}. Interesting studies about testing spacetimes within alternative theories of gravity with the EHT images have been presented in \cite{tests2020a,tests2020b,tests2021a,tests2021b,tests2021c,tests2022,vagnozzi2023,salehi2024,khodadi2024}. Several works analyze the prospects for improvements in the facilities used in the observation of black hole shadows \cite{ivanov2019,mikheeva2020,roelofs2019,fromm2021,doeleman2023,ayzenberg2025}. 

Black bounces, introduced a few years ago by Simpson and Visser \cite{simpsonvisser2019}, are spacetimes that interpolate between nonsingular black holes and wormholes, depending on the parameters of the particular model adopted. They present a minimal surface dubbed bounce, which is hidden behind an event horizon. These objects are free from singularities because the radial function exhibits a nontrivial bouncing behavior that keeps curvature invariants bounded and they allow congruences to reexpand beyond the throat to ensure geodesic completeness \cite{carballo2020,lobo2021}. Many articles have recently appeared in the literature \cite{bronnikov2022,rodrigues2023,fabris2023,alencar2024,junior2024a,junior2024b, atazadeh,pereira2025a,pereira2025b,lessaolmo2025,alencar2025,rois2025} exploring different sources or gravitational theories. Rotating black bounces were obtained \cite{mazza2021} by using the modified Newman-Janis procedure for the Simpson-Visser geometry. Black bounces have also appeared in studies of gravitational lensing \cite{islam2021,nascimento2020,cheng2021,ghosh2022}, photon rings, and shadows \cite{guerrero2021,pal2021,guo2022,pal2023,olmo2023,guerrero2022,dasilva2023}. Related works on shadows produced by wormholes have been published in recent years \cite{throat_1,shaikh2018,shaikh2019b,wang2020,throat_2,throat_3,throat_4,kar2025,elongated}.

In this article, we obtain rotating black bounces by applying a generalized Newman-Janis algorithm to a class of static solutions with an anisotropic fluid \cite{lessaolmo2025} and we study the shadows cast by them. In Sec. \ref{sec:rotating_black_bounces}, we find the rotating spacetimes and analyze their main features. In Sec. \ref{sec:light_propagation}, we derive the geodesics of photons and separate them in terms of the conserved quantities, which are used to calculate the contour of the shadow for an arbitrary value of the inclination angle of the observer. In Sec. \ref{sec:results}, we show some examples of shadows calculated analytically for relevant values of the model parameters and we compare them with those produced with a ray-tracing method \cite{pyhole}, considering  an equatorial observer. Finally, in Sec. \ref{sec:final_remarks}, we summarize and discuss our main results. Throughout the paper, we use the $(-,+,+,+)$ signature for the spacetime metric and adopt units such that $c=G=1$.

\section{Rotating black bounces with an anisotropic fluid}\label{sec:rotating_black_bounces}

We initially consider a static metric introduced by Lessa and Olmo \cite{lessaolmo2025}. It describes a spherically symmetric geometry generated by an
anisotropic fluid whose stress-energy tensor can be written as
\begin{equation}
    T_{\mu\nu}=(\rho+p_t)u_\mu u_\nu + p_t g_{\mu\nu}+ (p_r-p_t)v_\mu v_\nu ,
\end{equation}
where $\rho$, $p_r$, and $p_t$ denote the energy density, the radial pressure, and the tangential pressure of the fluid, respectively. The vector $u^\mu$ corresponds to the four-velocity of the fluid, normalized according to $g_{\mu\nu}u^\mu u^\nu=-1$, while $v_\mu$ is a spacelike unit vector with normalization $g_{\mu\nu}v^\mu v^\nu=+1$. For the case of our interest, we have $p_r=-\rho$ and $p_t=\omega \rho$, with $\omega$ being a dimensionless constant, and the energy density is given by
\begin{equation}
    \rho(r)=\frac{\rho_0}{(r^2+l^2)^{1+\omega}},
\end{equation}
where $\rho_0$ is a positive constant related to the energy density at the origin and $l$ is a parameter with dimensions of length, introduced to avoid a singularity at $r=0$ (which is motivated by the Simpson-Visser solution \cite{simpsonvisser2019}). Finally, this leads to the metric, in Schwarzschild coordinates,
\begin{equation}
    ds^2=-G(r)\,dt^2+\frac{dr^2}{F(r)}+H(r)\,d\Omega^2,
    \label{lessa_olmo_metric}
\end{equation}
where $d\Omega^2=d\theta^2+\sin^2\theta\,d\phi^2$ is the standard line element on a unit 2-sphere and the functions $G(r)$, $F(r)$, and $H(r)$ are given by
\begin{align}
    &G(r)=1-\frac{2m}{\sqrt{r^2+l^2}}+\frac{\rho_0}{(2\omega-1)(r^2+l^2)^\omega},\label{funcion_G} \\
    &F(r)=\left(1+\frac{l^2}{r^2}\right)G(r),\label{funcion_F} \\
    &H(r)=r^2+l^2, \label{funcion_H}
\end{align}
representing a black bounce (i.e. non-singular black hole or traversable wormhole, depending on the values of the parameters) with mass $m$ and throat radius $l$. Given the form of $\rho(r)$, we will restrict our attention to cases with $\omega>-1$, which lead to a decreasing energy density. Note also that the metric (\ref{lessa_olmo_metric}) is asymptotically flat if $\omega>0$ (with $\omega\neq1/2$). Therefore, we will focus exclusively on the cases with $\omega>0$. This includes the Reissner-Nordström metric, which is the particular case we get for $\omega=1$, $l=0$, and identifying $\rho_0$ with the square of the electric charge.

The rotating counterpart of the metric (\ref{lessa_olmo_metric}) is obtained by using the Azreg-A\"{i}nou procedure \cite{azreg2014a,azreg2014b}, which is a modification of the Newman-Janis algorithm \cite{newman65}. In Boyer–Lindquist coordinates it takes the form
\begin{equation}
    ds^2=\frac{\Psi}{\rho^2}\left[-\left(1-\frac{2f}{\rho^2}\right)dt^2+\frac{\rho^2}{\Delta}dr^2-\frac{4af\sin^2\theta}{\rho^2}dtd\phi+\rho^2d\theta^2+\frac{\Sigma\sin^2\theta}{\rho^2}d\phi^2\right],
    \label{rotating_lessa_olmo}
\end{equation}
where the parameter $a$ is the spin of the central object, and the stationary metric functions $\rho^2$, $f$, $\Delta$, and $\Sigma$ can be related to the static metric functions by
\begin{align}
    &\rho^2=\sqrt{\frac{F}{G}}H+a^2\cos^2\theta, \\
    &2f=\sqrt{\frac{F}{G}}H-FH, \\
    &\Delta=FH+a^2,\\
    &\Sigma=\left(\sqrt{\frac{F}{G}}H+a^2\right)^2-a^2\Delta\sin^2\theta,
\end{align}
acquiring the form
\begin{align}
    &\rho^2(r,\theta)=\frac{(r^2+l^2)^{3/2}}{|r|}+a^2\cos^2\theta,\\
    &2f(r)=\frac{(r^2+l^2)^{3/2}}{|r|}-\frac{(r^2+l^2)^2}{r^2}\left(1-\frac{2m}{\sqrt{r^2+l^2}}+\frac{\rho_0}{(2\omega-1)(r^2+l^2)^\omega}\right),\\
    &\Delta(r)=\frac{(r^2+l^2)^2}{r^2}\left(1-\frac{2m}{\sqrt{r^2+l^2}}+\frac{\rho_0}{(2\omega-1)(r^2+l^2)^\omega}\right)+a^2, \label{delta}\\
    &\Sigma(r,\theta)=\left[\frac{(r^2+l^2)^{3/2}}{|r|}+a^2\right]^2-a^2\Delta(r)\sin^2\theta.
\end{align}
The remaining function $\Psi(r,\theta)$ in (\ref{rotating_lessa_olmo}) can, in principle, be determined from the Einstein equations once the form of the fluid stress–energy tensor is specified (see \cite{azreg2014a} and \cite{azreg2014b}). Owing to its conformal factor character, its explicit form will not be relevant at this stage, and we shall return to it only when necessary later on.

Note that $r$ can take both positive and negative values, i.e., $r\in(-\infty,\infty)$. Furthermore, the metric (\ref{rotating_lessa_olmo}) is symmetric under the reflection $r\rightarrow-r$, implying that the spacetime it describes consists of two identical regions smoothly joined at $r=0$. Given this symmetry, from now on we will restrict our analysis to $r\geq0$ without loss of generality.

\begin{figure}[t]
    \centering
    \includegraphics[width=0.7\linewidth]{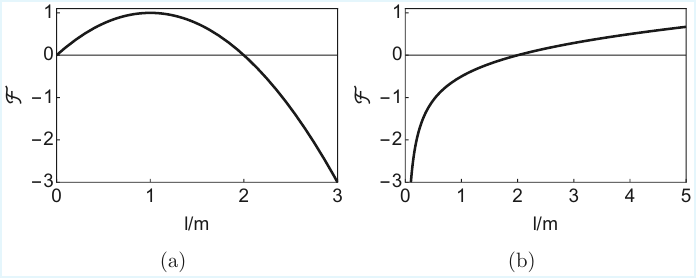}
    \begin{subfigure}{.4\linewidth}
        \phantomsubcaption\label{f_omega_1}
    \end{subfigure}
    \begin{subfigure}{.4\linewidth}
        \phantomsubcaption\label{f_omega_1_4}
    \end{subfigure}
    \caption{Plot of the function $\mathcal{F}$ with $m=1$: (\subref{f_omega_1}) $\omega=1$ and (\subref{f_omega_1_4}) $\omega=1/4$.}
    \label{fig:funcion_f}
\end{figure}

\begin{figure}[t]
    \centering
    \includegraphics[width=\linewidth]{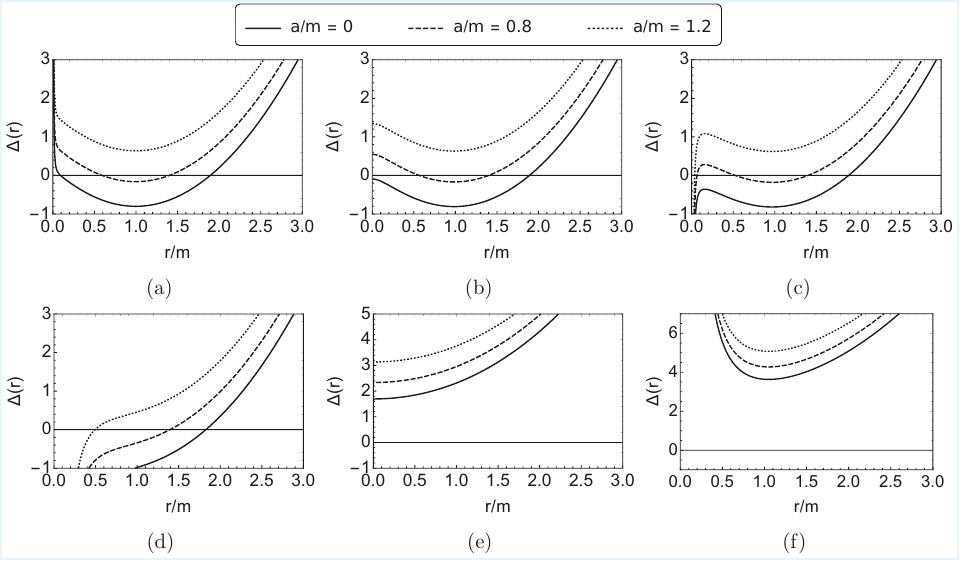}
    \begin{subfigure}{.32\linewidth}
        \phantomsubcaption\label{case1}
    \end{subfigure}
    \begin{subfigure}{.32\linewidth}
        \phantomsubcaption\label{case2}
    \end{subfigure}
    \begin{subfigure}{.32\linewidth}
        \phantomsubcaption\label{case3}
    \end{subfigure}
    \begin{subfigure}{.32\linewidth}
        \phantomsubcaption\label{case4}
    \end{subfigure}
    \begin{subfigure}{.32\linewidth}
        \phantomsubcaption\label{case5}
    \end{subfigure}
    \begin{subfigure}{.32\linewidth}
        \phantomsubcaption\label{case6}
    \end{subfigure}
    \caption{Plots of $\Delta(r)$ for the different cases, all of them with $\omega=m=1$ and $\rho_0=0.2$. (\subref{case1}) $l=0.05$ (case 1). (\subref{case2}) $\mathcal{F}(l\simeq0.106,\omega,m)=\rho_0$ (case 2). (\subref{case3}) $l=0.15$ (case 3). (\subref{case4}) $l=0.5$ (case 4). (\subref{case5}) $\mathcal{F}(l\simeq1.894,\omega,m)=\rho_0$ (case 5). (\subref{case6}) $l=2$ (case 6).  Note that $\omega=m=1$ leads to $\mathcal{F}_{max}=1$, with $l_{max}=1$.}
    \label{fig:cases}
\end{figure}

Having introduced the metric, we now turn to the analysis of its possible horizons. These are determined by the condition that the radial component of the inverse metric vanishes, namely $g^{rr}=\Delta/\Psi=0$. Assuming $\Psi$ to be a regular function, this happens when $\Delta=0$. Note that $\Delta \underset{r \to \infty}{\longrightarrow} + \infty$ in the expression (\ref{delta}). On the other hand, the behavior of $\Delta(r)$ near the origin can be seen from
\begin{equation}
    \Delta(r)=\frac{l^{4-2\omega}}{(2\omega-1)r^2}\Big[\rho_0-\mathcal{F}(l,\omega,m)\Big]+a^2+l^2-m\,l+\frac{(1-\omega)\rho_0\,l^{2-2\omega}}{(2\omega-1)}+\frac{l^{2-2\omega}}{(2\omega-1)}\Big[\rho_0-\mathcal{F}(l,\omega,m)\Big]+\mathcal{O}(r^2),\label{delta_en_cero}
\end{equation}
where the function $\mathcal{F}$ is given by
\begin{equation}
    \mathcal{F}(l,\omega,m)=(2\omega-1)l^{2\omega-1}(2m-l).
\end{equation}
Except for the case $\rho_0=\mathcal{F}$, Eq. (\ref{delta_en_cero}) gives us $\Delta \underset{r \to 0}{\longrightarrow} \pm\infty$, where the sign is determined by the sign of $(\rho_0-\mathcal{F})/(2\omega-1)$. So we have two different analyses, accordingly to the value of $\omega$: $0<\omega<1/2$ and $\omega>1/2$. Considering $\omega>1/2$ ($0<\omega<1/2$), if $\rho_0<\mathcal{F}$ ($\rho_0>\mathcal{F}$), Bolzano theorem tells us that $\Delta$ will have at least one root, a horizon, regardless of the value of $a$, while for $\mathcal{F}<\rho_0$ ($\mathcal{F}>\rho_0$), it may have no roots. Note that, for given $m$ and $\omega>1/2$, the function $\mathcal{F}$ has a maximum value, $\mathcal{F}_{max}$, given by
\begin{equation}
    \mathcal{F}_{max}=\left[\left(\frac{2\omega-1}{\omega}\right)m\right]^{2\omega},
\end{equation}
which occurs at $l=l_{max}=\left(\frac{2\omega-1}{\omega}\right)m$ (see Fig. \ref{f_omega_1}). Whereas for $0<\omega<1/2$, $\mathcal{F}$ is a monotonically increasing function of $l$ (Fig. \ref{f_omega_1_4}) and for all $\rho_0$, there exists a value $l_c$ for which $\mathcal{F}(l_c)=\rho_0$. This leads to six cases of interest:
\begin{enumerate}
    \item For $\mathcal{F}(l,\omega>1/2,m)<\rho_0\leq\mathcal{F}_{max}$ with $l<l_{max}$, we have a (two-way) traversable wormhole (no horizons) or a regular black hole\footnote{The wormhole's throat is inside an event horizon.} with one or two horizons depending on $a$ (Fig. \ref{case1}).
    \item For $\rho_0=\mathcal{F}(l,\omega>1/2,m)$ and $l\leq l_{max}$, we have a (two-way) traversable wormhole (no horizons) or a regular black hole with one or two  horizons depending on $a$ (Fig. \ref{case2}).\footnote{Strictly, for the case $\rho_0=F(l_{max},\omega>1/2,m)$ and $a=0$ is a one-way traversable wormhole because it has only one horizon but it is on the throat.}
    \item In some cases, for $\rho_0\lesssim\mathcal{F}(l,\omega>1/2,m)$ with $l\lnsim l_{max}$ or $\rho_0>\mathcal{F}(l,0<\omega<1/2,m)$ with $0<l\ll l_c$, we have a regular black hole with one, two, or three  horizons depending on $a$ (Fig. \ref{case3}).
    \item If we do not fall into special case 3, for $\rho_0<\mathcal{F}(l,\omega>1/2,m)$, or $\rho_0\gtrsim \mathcal{F}(l,0<\omega<1/2,m)$, we have a regular black hole with only one horizon, regardless of the value of $a$ (Fig. \ref{case4}).
    \item For $\rho_0=\mathcal{F}(l,\omega>1/2,m)$ with $l>l_{max}$ or $\rho_0=\mathcal{F}(l,0<\omega<1/2,m)$, there are no horizons and we have a (two-way) traversable wormhole, regardless of the value of $a$ (Fig. \ref{case5}).
    \item For $\rho_0>\mathcal{F}_{max}$ with $\omega>1/2$ or $\mathcal{F}(l,\omega>1/2,m)<\rho_0\leq\mathcal{F}_{max} \text{ with }\,l>l_{max}$ or $\rho_0<\mathcal{F}(l,0<\omega<1/2,m)$, there are no horizons and we have a (two-way) traversable wormhole, regardless of the value of $a$ (Fig. \ref{case6}).
\end{enumerate}

\begin{figure}[t]
    \centering
    \includegraphics[width=.7\linewidth]{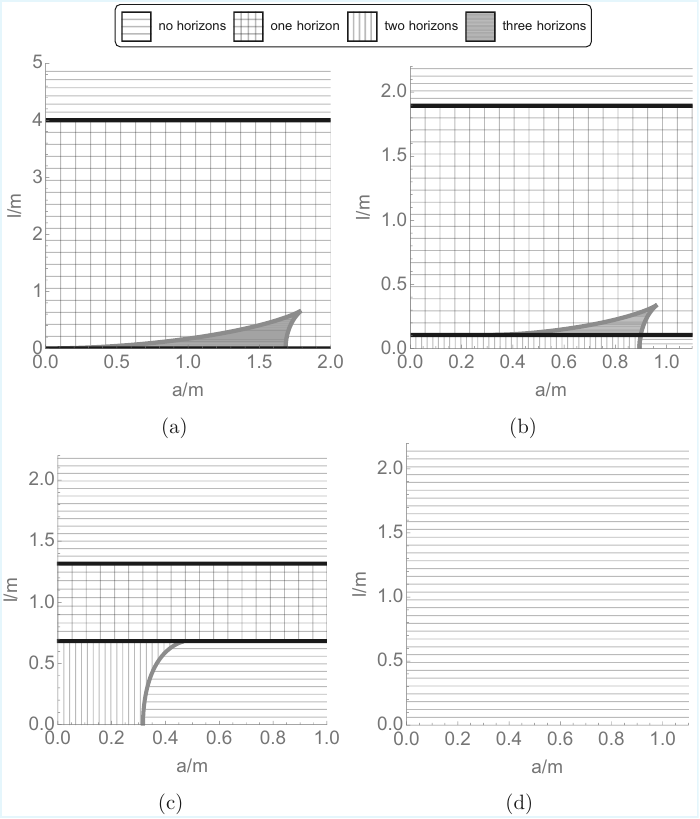}
    \begin{subfigure}{.35\linewidth}
        \phantomsubcaption\label{omega_1_4}
    \end{subfigure}
    \begin{subfigure}{.35\linewidth}
        \phantomsubcaption\label{rho02}
    \end{subfigure}
    \begin{subfigure}{.35\linewidth}
        \phantomsubcaption\label{rho09}
    \end{subfigure}
    \begin{subfigure}{.35\linewidth}
        \phantomsubcaption\label{rho12}
    \end{subfigure}
    \caption{Regions with different horizon structures in the parameter space $(a,l)$ for different cases, all of them with $m=1$. (\subref{omega_1_4}) $\omega=1/4$, $\rho_0=0.5$. (\subref{rho02}) $\omega=1$, $\rho_0=0.2$. (\subref{rho09}) $\omega=1$, $\rho_0=0.9$. (\subref{rho12}) $\omega=1$, $\rho_0=1.2$. The black lines indicate the cases where $\rho_0=\mathcal{F}(l,\omega,m)$.}
    \label{fig:phase_diagrams}
\end{figure}

We can summarize the spectrum of possible cases using ``phase diagrams" in the parameter space $(a,l)$, as shown in Fig. \ref{fig:phase_diagrams}. In Fig. \ref{rho12} we have no horizons because $\rho_0>\mathcal{F}_{max}$ with $\omega>1/2$ and we are always in case 6. On the other hand, for Figs. \ref{rho02} and \ref{rho09}, if we analyze from smaller values of $l$, we start with $\mathcal{F}(l,\omega>1/2,m)<\rho_0<\mathcal{F}_{max}$ and $l<l_{max}$, corresponding to case 1, where we have either one or zero horizons depending on $a$. As $l$ increases, we transition to case 2 (indicated by the lower black line in Fig. \ref{fig:phase_diagrams}), where $\mathcal{F}(l,\omega>1/2,m)=\rho_0$ and $l<l_{max}$. From there, we may first reach case 3 (Fig. \ref{rho02}), where up to three horizons are possible, or directly arrive at case 4 (Fig. \ref{rho09}), with exactly one horizon, depending on $\rho_0$. In both cases, $\mathcal{F}(l,\omega>1/2,m)>\rho_0$. For Fig. \ref{omega_1_4}, with $0<\omega<1/2$, we start here, first with case 3 and then case 4. Therefore, as shown in the graph of $\mathcal{F}(l,\omega>1/2,m)$ in Fig. \ref{f_omega_1}, we again encounter $\mathcal{F}=\rho_0$ (indicated by the upper black line in Fig. \ref{fig:phase_diagrams}), but now with $l>l_{max}$ (or directly, $\rho_0=\mathcal{F}$ with $0<\omega<1/2$), placing us in case 5. Finally, as $l$ continues to increase, $\mathcal{F}(l,\omega>1/2,m)$ becomes smaller than $\rho_0$ while still in the regime $l>l_{max}$ (or $\rho_0<\mathcal{F}$ with $0<\omega<1/2$), and we transition to case 6, where horizons do not exist.

\section{Light propagation}\label{sec:light_propagation}

Assuming the separability of the problem, the Jacobi action $S$ for the propagation of light is given by
\begin{equation}
    S=-E\,t+L_z\,\phi+S_r(r)+S_\theta(\theta),
\end{equation}
where the conserved quantities $E$ and $L_z$ are the energy and the component of the angular momentum along the axis of rotation, respectively. The Hamilton-Jacobi equation reads
\begin{equation}
    g^{\mu\nu}\frac{\partial S}{\partial x^\mu}\frac{\partial S}{\partial x^\nu}=0,
\end{equation}
which results in
\begin{equation}
    -\frac{\Sigma}{\Psi\Delta}E^2+\frac{4fa}{\Psi\Delta}E L_z+\frac{\Delta-a^2\sin^2\theta}{\Psi\Delta\sin^2\theta}{L_z}^2+\frac{\Delta}{\Psi}\left(\frac{dS_r}{dr}\right)^2+\frac{1}{\Psi}\left(\frac{dS_\theta}{d\theta}\right)^2=0\,.
\end{equation}
So we get
\begin{equation}
    -\left[\frac{\left[(r^2+l^2)^{3/2}+a^2 r\right]^2}{\Delta r^2}-a^2\sin^2\theta\right]E^2+\frac{4fa}{\Delta}E L_z+\left(\frac{1}{\sin^2\theta}-\frac{a^2}{\Delta}\right){L_z}^2+\Delta\left(\frac{dS_r}{dr}\right)^2+\left(\frac{dS_\theta}{d\theta}\right)^2=0,
\end{equation}
which can be written more transparently as
\begin{equation}
    \frac{1}{\Delta(r)}\left[\left(\frac{(r^2+l^2)^{3/2}}{r}+a^2\right)E-a L_z\right]^2-\Delta(r) \left(\frac{dS_r}{dr}\right)^2=\left(\frac{dS_\theta}{d\theta}\right)^2+\left(Ea\sin\theta-\frac{L_z}{\sin\theta}\right)^2\,.
\end{equation}
The equation above allows us to define a separation constant\footnote{It is known as the Carter constant \cite{carter68}, associated with a fourth conserved quantity (the other three are the photon null mass, $E$, and $L_z$).} $K$ and give us the equations for $S_r$ and $S_\theta$
\begin{align}
     &K=\frac{1}{\Delta(r)}\left[\left(\frac{(r^2+l^2)^{3/2}}{r}+a^2\right)E-a L_z\right]^2-\Delta(r) \left(\frac{dS_r}{dr}\right)^2, \label{eq_s_r}\\
     &K=\left(\frac{dS_\theta}{d\theta}\right)^2+\left(Ea\sin\theta-\frac{L_z}{\sin\theta}\right)^2. \label{eq_s_theta}
\end{align}
Hence, we have
\begin{align}
    &S_r=\pm\int^r\frac{\sqrt{R}}{\Delta}dr,\\
    &S_\theta=\pm\int^\theta\sqrt{\Theta}\,d\theta,
\end{align}
where
\begin{align}
    &R=\left[\left(\frac{(r^2+l^2)^{3/2}}{r}+a^2\right)E-a L_z\right]^2-\Delta K, \label{def_R}\\
    &\Theta=K-\left(Ea\sin\theta-\frac{L_z}{\sin\theta}\right)^2. \label{def_theta}
\end{align}
With the aim of finding the equations of motion, we note that the relation between the Jacobi action and the momentum $p^\mu$ is
\begin{equation}
    \frac{\partial S}{\partial x^\mu}=p_\mu.
\end{equation}
Therefore, we get
\begin{equation}
    \frac{d x^\mu}{d\lambda}=p^\mu=g^{\mu\nu}p_\nu=g^{\mu\nu}\frac{\partial S}{\partial x^\nu},
\end{equation}
where $\lambda$ is an affine parameter. From the definitions in Eqs. (\ref{def_R}) and (\ref{def_theta}), and using Eqs. (\ref{eq_s_r}) and (\ref{eq_s_theta}) we obtain
\begin{align}
    &R=\Delta^2\left(\frac{dS_r}{dr}\right)^2=\Delta^2\left(\frac{1}{g^{rr}}\frac{d r}{d\lambda}\right)^2=\Psi^2\left(\frac{d r}{d\lambda}\right)^2, \label{eom1}\\
    &\Theta=\left(\frac{dS_\theta}{d\theta}\right)^2=\left(\frac{1}{g^{\theta\theta}}\frac{d \theta}{d\lambda}\right)^2=\Psi^2\left(\frac{d \theta}{d\lambda}\right)^2. \label{eom2}
\end{align}
And for $t$ and $\phi$ we have
\begin{align}
    &\frac{d t}{d\lambda}=g^{tt}\frac{\partial S}{\partial t}+g^{t\phi}\frac{\partial S}{\partial \phi}=\frac{\Sigma}{\Psi\Delta}E-\frac{2fa}{\Psi\Delta}L_z,\\
    &\frac{d \phi}{d\lambda}=g^{\phi t}\frac{\partial S}{\partial t}+g^{\phi\phi}\frac{\partial S}{\partial \phi}=-\frac{2fa}{\Psi\Delta}E+\frac{\Delta-a^2\sin^2\theta}{\Psi\Delta\sin^2\theta}L_z,
\end{align}
or more explicitly,
\begin{align}
    &\Psi\frac{d t}{d\lambda}=\frac{1}{\Delta}\left[\frac{(r^2+l^2)^{3/2}}{r}+a^2\right]\left[\left(\frac{(r^2+l^2)^{3/2}}{r}+a^2\right)E-a L_z\right]-a(a\sin^2\theta E-L_z), \label{eom3}\\
    &\Psi\frac{d \phi}{d\lambda}=-\frac{a}{\Delta}\left[\left(\frac{(r^2+l^2)^{3/2}}{r}+a^2\right)E+a L_z\right]+aE+\frac{L_z}{\sin^2\theta}. \label{eom4}
\end{align}
Defining $\xi=L_z/E$ and $\eta=Q/E^2$, with $Q=K-(Ea-L_z)^2$ (so $K/E^2=\eta+(a-\xi)^2$), we get
\begin{align}
    &\mathcal{R}=\frac{R}{E^2}=\left[\frac{(r^2+l^2)^{3/2}}{r}+a^2-a \xi\right]^2-\Delta \left[\eta+(a-\xi)^2\right], \label{R_function} \\
    &\vartheta=\frac{\Theta}{E^2}=\eta+\cos^2\theta\left(a^2-\frac{\xi^2}{\sin^2\theta}\right),
\end{align}
and the equations of motion (\ref{eom1}), (\ref{eom2}), (\ref{eom3}), and (\ref{eom4}) can be rewritten as follows:
\begin{align}
    &\Psi\frac{dr}{d\tilde{\lambda}}=\pm\sqrt{\mathcal{R}},\label{eq_r}\\
    &\Psi\frac{d\theta}{d\tilde{\lambda}}=\pm\sqrt{\vartheta},\label{eq_theta}\\
    &\Psi\frac{dt}{d\tilde{\lambda}}=\frac{1}{\Delta}\left[\frac{(r^2+l^2)^{3/2}}{r}+a^2\right]\left[\frac{(r^2+l^2)^{3/2}}{r}+a^2-a \xi\right]-a(a\sin^2\theta-\xi),\label{eq_t}\\
    &\Psi\frac{d\phi}{d\tilde{\lambda}}=-\frac{a}{\Delta}\left[\frac{(r^2+l^2)^{3/2}}{r}+a^2+a \xi\right]+a+\frac{\xi}{\sin^2\theta},\label{eq_phi}
\end{align}
where $\tilde{\lambda}=E\lambda$.

\subsection{Impact parameters}\label{sec:impact}

For an observer located at the point $(r_0,\, \theta_0,\, \phi_0=0)$, the impact parameters for a given light ray (defined by $\xi$ and $\eta$) are
\begin{align}
    &x=-{r_0}^2\sin\theta_0\left.\frac{d\phi}{dr}\right|_{(r_0,\theta_0,0)},\\
    &y={r_0}^2\left.\frac{d\theta}{dr}\right|_{(r_0,\theta_0,0)}.
\end{align}
Using Eqs. (\ref{eq_r}), (\ref{eq_theta}), and (\ref{eq_phi}) we get
{\small 
\begin{align}
    &\frac{d\phi}{dr}=\pm\left\{-\frac{a}{\Delta}\left[\frac{(r^2+l^2)^{3/2}}{r}+a^2+a \xi\right]+a+\frac{\xi}{\sin^2\theta}\right\}\left\{\left[\frac{(r^2+l^2)^{3/2}}{r}+a^2-a \xi\right]^2-\Delta \left[\eta+(a-\xi)^2\right]\right\}^{-1/2},\\
    &\frac{d\theta}{dr}=\pm\sqrt{\vartheta}\left\{\left[\frac{(r^2+l^2)^{3/2}}{r}+a^2-a \xi\right]^2-\Delta \left[\eta+(a-\xi)^2\right]\right\}^{-1/2}.
\end{align}
}

As mentioned above, if we consider $\omega>0$ the spacetime is asymptotically flat, so for a distant observer we have
\begin{align}
    &x=\lim_{r_0\to\infty}-{r_0}^2\sin\theta_0\left.\frac{d\phi}{dr}\right|_{(r_0,\theta_0,0)}=-\frac{\xi}{\sin\theta_0},\label{param_x_theta0}\\
    &y=\lim_{r_0\to\infty}{r_0}^2\left.\frac{d\theta}{dr}\right|_{(r_0,\theta_0,0)}=\pm\sqrt{\eta+a^2\cos^2\theta_0-\frac{\xi^2}{\tan^2\theta_0}},\label{param_y_theta0}
\end{align}
where we have used that $\Delta(r_0)\sim {r_0}^2$ at large distances $r_0$. We also disregard the $\pm$ sign in Eq. (\ref{param_x_theta0}) as it is irrelevant for constructing a silhouette when considering all possible values of $\xi$ (both positive and negative). This is not the case for Eq. (\ref{param_y_theta0}) where the $\pm$ sign is necessary to ensure that a silhouette would be symmetric with respect to the plane $z=0$, in accordance with the symmetry of the spacetime. Note that, when the observer is situated in the equatorial plane of the black hole, the inclination angle is $\theta_0=\pi/2$, and we get
\begin{align}
    &x=-\xi,\label{param_x}\\
    &y=\pm\sqrt{\eta}.\label{param_y}
\end{align}
Another particular situation of interest corresponds to a polar observer, in which $\theta_0=0,\,\pi$. By squaring Eqs. (\ref{param_x_theta0}) and (\ref{param_y_theta0}), after some algebra, we obtain the relation
\begin{equation}
    x^2+y^2=\eta+a^2.
\end{equation}
The condition $\xi=0$ always holds in this case.

\subsection{Shadow}\label{sec:shadow}

The corresponding shadow observed at large distances is characterized by its boundary, which corresponds to the limit trajectories of photons (originate at infinity) that separate the two possible behaviors: to escape to infinity again or to be captured\footnote{We are not considering light that could come from the other side of the throat, i.e., from the region with $r<0$.}. From Eq. (\ref{eq_r}), the turning point of a photon trajectory, if it exists, is given by the largest root of the effective potential $V_{eff}=-\mathcal{R}/\Psi^2$, because movement is only possible in the region $V_{eff}<0$. Assuming $\Psi$ to be a regular function, this would occur at the greatest zero of $\mathcal{R}$. The critical point occurs when there are no more roots and the photons continue to $r=0$, the throat, and never return. The function $V_{eff}$ is well behaved in the region $(0,+\infty)$, so the only way roots can disappear in that region is when a maximum of $V_{eff}$ crosses per zero (which gives an unstable orbit with constant $r$). However, there is another possibility. If we have no horizons, photons can even return from $r=0$. However, a root at $r=0$ can appear or disappear abruptly, depending on the value of the limit $V_{eff}(r\to0)$, which could be $+\infty$, $0$, or $-\infty$ (we will see this in more detail below). 

Let us first focus on the more familiar case of unstable orbits with constant $r$, which form what is known as the photon sphere. For these trajectories, and always assuming $\Psi$ to be a regular function, we have the conditions $\mathcal{R}=0$, $d\mathcal{R}/dr=0$, and $d^2\mathcal{R}/dr^2>0$. From equation $\mathcal{R}=0$ we have
\begin{equation}
    \eta=\frac{1}{\Delta}\left[\frac{(r^2+l^2)^{3/2}}{r}+a^2-a \xi\right]^2-(a-\xi)^2\,.
\end{equation}
Replacing this in equation $d\mathcal{R}/dr=0$ we arrive to
\begin{equation}
    2\left[\frac{(r^2+l^2)^{3/2}}{r}+a^2-a \xi\right]\left[3\sqrt{r^2+l^2}-\frac{(r^2+l^2)^{3/2}}{r^2}\right]-\frac{{\Delta^\prime}}{\Delta}\left[\frac{(r^2+l^2)^{3/2}}{r}+a^2-a \xi\right]^2=0,\label{eq_shadow}
\end{equation}
which gives two pairs of solutions. In one case,
\begin{equation}
    \left[\frac{(r^2+l^2)^{3/2}}{r}+a^2-a \xi\right]=0,
\end{equation}
and hence
\begin{align}
    &\xi=\frac{(r^2+l^2)^{3/2}}{a\,r}+a,\\
    &\eta=-\frac{(r^2+l^2)^{3}}{a^2 r^2}.
\end{align}
This is not a physically meaningful solution because it gives $y^2<0$ in Eq. (\ref{param_y_theta0}). Therefore, the relevant solution of Eq. (\ref{eq_shadow}) is the other one, namely
\begin{align}
    &\xi=\frac{(r^2+l^2)^{3/2}}{a\,r}+a-\frac{2\Delta}{a\Delta^\prime}\left[3\sqrt{r^2+l^2}-\frac{(r^2+l^2)^{3/2}}{r^2}\right],\label{eq_xhi}\\
    &\eta=\frac{4\Delta}{(\Delta^\prime)^2}\left[3\sqrt{r^2+l^2}-\frac{(r^2+l^2)^{3/2}}{r^2}\right]^2-\left[\frac{(r^2+l^2)^{3/2}}{a\,r}-\frac{2\Delta}{a\Delta^\prime}\left[3\sqrt{r^2+l^2}-\frac{(r^2+l^2)^{3/2}}{r^2}\right]\right]^2. \label{eq_eta}
\end{align}
The last step to obtain the contour of the shadow is to find the values of $r$ for which we have the unstable spherical orbits. On the one hand, we have $\mathcal{R}=d\mathcal{R}/dr=0$ for the spherical orbits, which are satisfied when $\xi$ and $\eta$ are given by Eqs. (\ref{eq_xhi}) and (\ref{eq_eta}) for the values of $r$ that give $y^2\geq0$ in Eq. (\ref{param_y_theta0}). On the other hand, unstable photon orbits occur when $d^2\mathcal{R}/dr^2>0$, according to Eq. (\ref{eq_r}). Choosing the values of $r$ that match both conditions, we find the shadow boundary using Eqs. (\ref{param_x_theta0}) and (\ref{param_y_theta0}) with $\xi$ and $\eta$ given by Eqs. (\ref{eq_xhi}) and (\ref{eq_eta}).

For the static and spherically symmetric case ($a=0$), described by the metric in Eq. (\ref{lessa_olmo_metric}), the shadow is circular, so its characterization reduces to determining the radius. From Eq. (\ref{R_function}), the effective potential takes the form of
\begin{equation}
    V_{eff}^{static}=-\frac{(r^2+l^2)^3}{r^2}\left(1-G(r)\frac{\eta+\xi^2}{r^2+l^2}\right)\,,
\end{equation}
and it has a root for
\begin{equation}
    \eta+\xi^2=\frac{r^2+l^2}{G(r)}.
\end{equation}
Replacing this into equation $d\mathcal{R}/dr=0$,
\begin{equation}
    \frac{G'(r_{ph})}{G(r_{ph})}=\frac{2r_{ph}}{r_{ph}^2+l^2},
\end{equation}
where $r_{ph}$ is the value of $r$ in the circular orbit of the photons. This leads to
\begin{equation}
    1-\frac{3m}{\sqrt{r_{ph}^2+l^2}}+\frac{(\omega+1)\rho_0}{(2\omega+1)(r_{ph}^2+l^2)^{\omega}}=0,
    \label{eq_x_ph_static}
\end{equation}
from which we can obtain $r_{ph}$, and the radius of the shadow $r_{sh}$ in the celestial sphere is (Eq. (29) of \cite{perlick2022})
\begin{equation}
    r_{sh}=\sqrt{x^2+y^2}=\sqrt{\eta+\xi^2}=\sqrt{\frac{r_{ph}^2+l^2}{G(r_{ph})}}.
    \label{eq_sh_static}
\end{equation}
For example, for $m=\omega=1$, from Eq. (\ref{eq_x_ph_static}) it follows\footnote{Equation (\ref{eq_x_ph_static}) is quadratic in this case, but only one of the two solutions gives an unstable orbit.}
\begin{equation}
    \sqrt{r_{ph}^2+l^2} = \frac{3+\sqrt{9-8\rho_0}}{2},
    \label{static_shadow_example}
\end{equation}
which, in principle, suggests that we do not have a shadow for $\rho_0>9/8$. We will return to this point after discussing the role of the throat.

Let us now examine the cases without horizons, where there could be a turning point even from $r=0$. As discussed in Sec. \ref{sec:rotating_black_bounces}, these correspond to cases 1, 2 (both depending on the value of $a$), 5, and 6. From Eq. (\ref{R_function}) one obtains
\begin{equation}
    \mathcal{R}(r\to 0)\simeq\frac{l^6}{r^2}-\Delta(r\to 0)\left[\eta+(a-\xi)^2\right]. \label{R_function_r_0}
\end{equation}
In cases 2 and 5, where $\rho_0=\mathcal{F}$, Eq. (\ref{delta_en_cero}) implies
\begin{equation}
    \Delta(r\to 0)\simeq a^2 + l^2 \left[1-\frac{m}{l}+\frac{(1-\omega)\rho_0}{(2\omega-1)l^{2\omega}}\right].
\end{equation}
Therefore, in these cases, we always have $\lim\limits_{r\to 0}\mathcal{R}=+\infty$ and $\lim\limits_{r\to 0}V_{eff}=-\infty$, whatever impact parameters we use. So for cases 2 and 5 we do not have any new possible contribution to the shadow (we never have a turning point at $r=0$).
For cases 1 and 6, the situation is different, because now $\rho_0\neq\mathcal{F}$. Then Eq. (\ref{delta_en_cero}) gives 
\begin{equation}
    \Delta(r\to 0)\simeq \frac{l^{4-2\omega}}{(2\omega-1)r^2}(\rho_0-\mathcal{F}),
\end{equation}
and Eq. (\ref{R_function_r_0}) leads to
\begin{equation}
    \mathcal{R}(r\to 0)\simeq \frac{(2\omega-1)l^6-l^{4-2\omega}(\rho_0-\mathcal{F})\left[\eta+(a-\xi)^2\right]}{(2\omega-1)r^2},
\end{equation}
which implies a critical turning point (a root of $\mathcal{R}$) at $r=0$ for
\begin{equation}
    \eta+(a-\xi)^2=\frac{(2\omega-1)l^{2+2\omega}}{\rho_0-\mathcal{F}}, \label{new_shadow}
\end{equation}
and, using Eqs. (\ref{param_x_theta0}) and (\ref{param_y_theta0}), results in
\begin{equation}
    y^2+(x+a\,\sin\theta_0)^2=\frac{(2\omega-1)l^{2+2\omega}}{\rho_0-\mathcal{F}}, \label{new_shadow_xy}
\end{equation}
that is, a circle.

\begin{figure}[t]
    \centering
    \includegraphics[width=.7\linewidth]{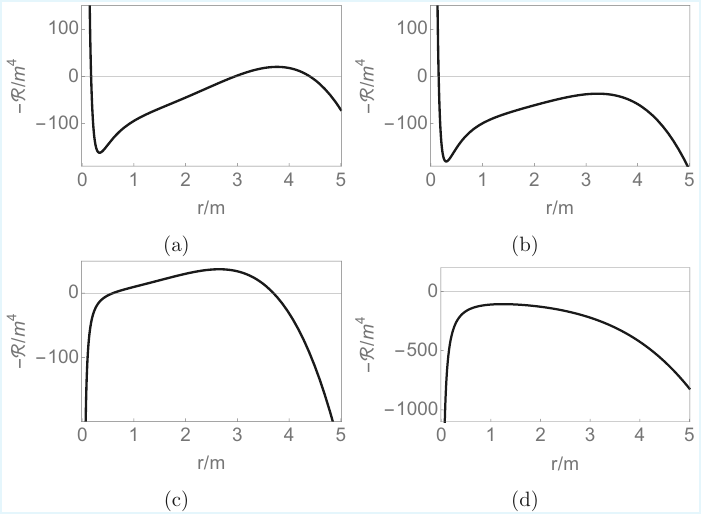}
    \begin{subfigure}{.45\linewidth}
        \phantomsubcaption\label{pot1}
    \end{subfigure}
    \begin{subfigure}{.45\linewidth}
        \phantomsubcaption\label{pot2}
    \end{subfigure}
    \begin{subfigure}{.45\linewidth}
        \phantomsubcaption\label{pot4}
    \end{subfigure}
    \begin{subfigure}{.45\linewidth}
        \phantomsubcaption\label{pot3}
    \end{subfigure}
    \caption{Different possibilities for the plots of $V_{eff}$ in the cases 1 without horizons and 6. In all plots $\omega=m=1$, $\rho_0=0.9$, $l=1.5$, and $\eta=0$. (\subref{pot1}) $a=2$ and $\xi=-7.8$; we start with a maximum with a positive value and $\lim\limits_{r\to 0}V_{eff}=+\infty$, the turning point is given by the highest root. (\subref{pot2}) $a=2$ and $\xi=-7.3$; now the maximum is negative but we still have $\lim\limits_{r\to 0}V_{eff}=+\infty$, so we still have a return (no shadow yet). (\subref{pot4}) $a=0.2$ and $\xi=-5.5$; the other possibility is that we have a positive maximum but $\lim\limits_{r\to 0}V_{eff}=-\infty$, so $r=0$ has no contribution to the shadow boundary. (\subref{pot3}) $a=2$ and $\xi=-3.8$; finally we have $\lim\limits_{r\to 0}V_{eff}=-\infty$ and a negative maximum, so this light ray is inside the shadow.}
    \label{fig:pot}
\end{figure}

\begin{figure}
    \centering
    \includegraphics[width=\linewidth]{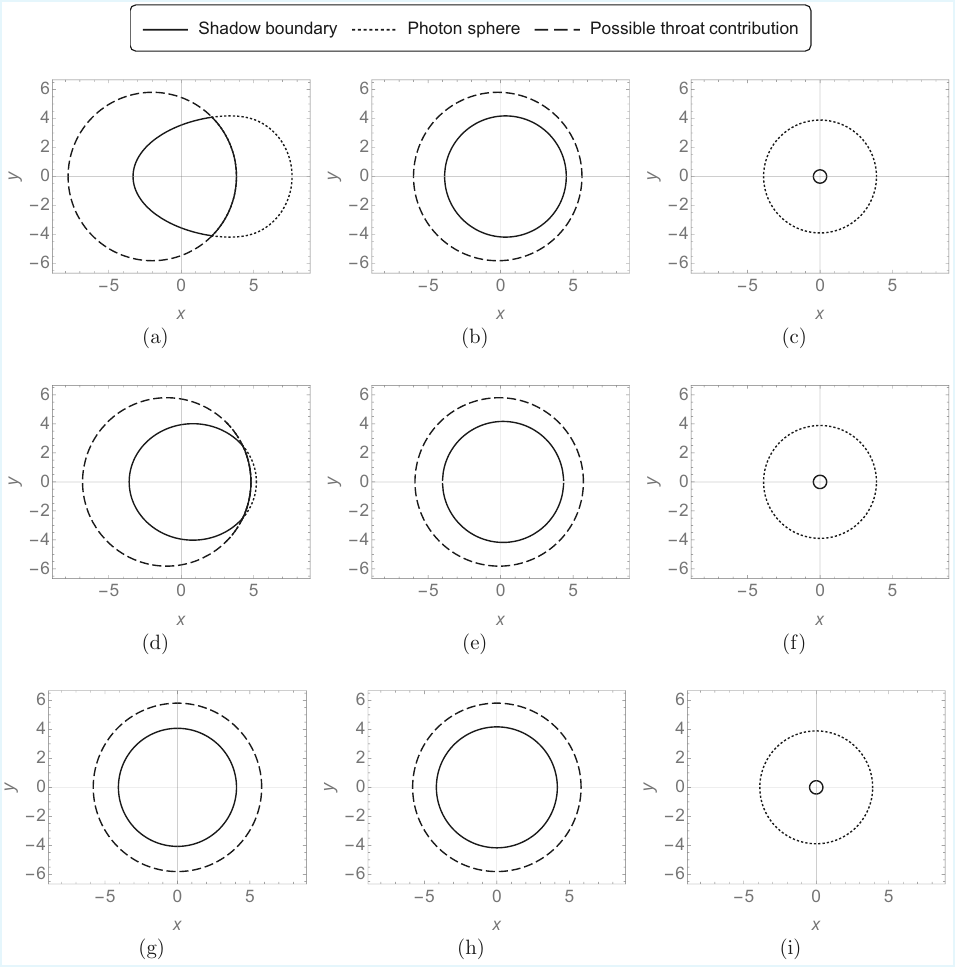}
    \begin{subfigure}{.32\linewidth}
        \phantomsubcaption\label{shadow_both}
    \end{subfigure}
    \begin{subfigure}{.32\linewidth}
        \phantomsubcaption\label{shadow_photon_sphere}
    \end{subfigure}
    \begin{subfigure}{.32\linewidth}
        \phantomsubcaption\label{shadow_throat}
    \end{subfigure}
    \begin{subfigure}{.32\linewidth}
        \phantomsubcaption\label{shadow_both_30}
    \end{subfigure}
    \begin{subfigure}{.32\linewidth}
        \phantomsubcaption\label{shadow_photon_sphere_30}
    \end{subfigure}
    \begin{subfigure}{.32\linewidth}
        \phantomsubcaption\label{shadow_throat_30}
    \end{subfigure}
    \begin{subfigure}{.32\linewidth}
        \phantomsubcaption\label{shadow_both_0}
    \end{subfigure}
    \begin{subfigure}{.32\linewidth}
        \phantomsubcaption\label{shadow_photon_sphere_0}
    \end{subfigure}
    \begin{subfigure}{.32\linewidth}
        \phantomsubcaption\label{shadow_throat_0}
    \end{subfigure}
    \caption{Examples of shadows arising in cases 1 without horizons and 6, produced by the two possible mechanisms: the photon sphere (outside the throat) and the wormhole throat. All plots correspond to $\omega=m=1$. The rows show different observer inclination angles: $\theta_0=\pi/2$ (top), $\theta_0=\pi/6$ (middle), and $\theta_0=0$ (bottom). The three columns correspond to different parameter choices: $\rho_0=0.9$, $l=1.5$, $a=2$ (left); $\rho_0=0.9$, $l=1.5$, $a=0.2$ (center); $\rho_0=1.05$, $l=0.5$, $a=0$ (right). See text for a detailed discussion.}
    \label{fig:shadows_examples}
\end{figure}

Therefore, for case 1 without horizons and case 6 (which never has horizons), we could have two different contributions to the shadow: the one due to the photon sphere, given by Eqs. (\ref{eq_xhi}) and (\ref{eq_eta}), and the one due to the wormhole throat, given by Eq. (\ref{new_shadow}). The possibility that the throat of a wormhole contributes to the shadow has already been discussed in the literature. However, in most cases this contribution arises from the presence of a photon sphere at the throat \cite{throat_1, shaikh2018, shaikh2019b, throat_2, throat_3, throat_4}. In contrast, in our case a root of the effective potential at the throat is the reason why we could have a throat contribution to the shadow, but there are no unstable spherical orbits there. To the best of our knowledge, only a few works in the bibliography report this kind of behavior, whether through turning points at a surface of horizonless compact objects \cite{pal2021} or through turning points at the singularity in the case of naked singularities \cite{joshi_2020,pal2023}.

To determine the final form of the shadow, given the two contributions, we need to choose the innermost closed curve that we can form with the two curves. To understand this, we can think that we are changing the impact parameters in order that the minimum value of $r$ in the light trajectory is moving closer and closer to the origin. The effective potential would take a form similar to Fig. \ref{pot1}, with a maximum with a positive value and $\lim\limits_{r\to 0}V_{eff}=+\infty$. Here, the turning point is given by the highest root of $V_{eff}$. If the maximum crosses the axis before we have had a turning point at $r=0$ (Fig. \ref{pot2}), then we will still have a root of $V_{eff}$ and light would escape. Finally, we will have $\lim\limits_{r\to 0}V_{eff}=-\infty$ (after the root at zero) and $V_{eff}$ will lose all their roots (Fig. \ref{pot3}), so light will be captured. Therefore, in this case, the shadow boundary will be given by the corresponding value in Eq. (\ref{new_shadow_xy}). However, if we have $\lim\limits_{r\to 0}V_{eff}=-\infty$ before the maximum of $V_{eff}$ crosses per zero (Fig. \ref{pot4}), then the throat will not contribute to the shadow in this direction (the one we took moving the impact parameters). In such a case, the shadow is given by Eqs. (\ref{eq_xhi}) and (\ref{eq_eta}).

Let us now return to the static case given by Eq. (\ref{static_shadow_example}); for example, for $l<l_{max}$. Note that for $\rho_0<\mathcal{F}_{max}=1$ we always have horizons because we are in (static) cases 1, 2, 3, or 4 and the shadow boundary is given by the photon sphere contribution. However, for $\rho_0>\mathcal{F}_{max}=1$, we are in case 6 (without horizons), so we could now have a throat contribution. If, for example, we take $l=0.5$ and $\rho_0=1.05$, Eqs. (\ref{eq_sh_static}) and (\ref{static_shadow_example}) give $r_{sh}\simeq 3.89$ and Eq. (\ref{new_shadow_xy}) gives $r_{sh}\simeq 0.46$, so the shadow is only formed by the throat contribution even before we reach the case $\rho_0=9/8$, from which we no longer have a photon sphere.

In Fig. \ref{fig:shadows_examples} we exhibit some illustrative examples of the different types of shadows that can arise from the two possible contributions: one corresponding to the photon sphere and the other to the throat. Each plot displays the dotted curve associated with the photon sphere—computed from Eqs. (\ref{param_x_theta0}) and (\ref{param_y_theta0}) using the impact parameters $\xi$ and $\eta$ given by Eqs. (\ref{eq_xhi}) and (\ref{eq_eta})—as well as the dashed curve associated with the throat, obtained from the root of $V_{eff}$ at $r=0$ and expressed in Eq. (\ref{new_shadow_xy}). The solid line, which always overlaps totally or partially with one of the other curves, represents the actual shadow boundary, which is the innermost closed curve formed with those two. Consequently, the dashed and dotted lines do not necessarily correspond to visible shadow features; they simply delineate the possible contributions from each mechanism. In Figs. \ref{shadow_photon_sphere}, \ref{shadow_photon_sphere_30}, \ref{shadow_both_0}, and \ref{shadow_photon_sphere_0}, the contour of the shadow is determined only by the photon sphere; in Figs. \ref{shadow_throat}, \ref{shadow_throat_30}, and \ref{shadow_throat_0} only by the throat contribution; while in Figs. \ref{shadow_both} and \ref{shadow_both_30} by a combination of both. 

Figure \ref{fig:shadows_examples} also illustrates the role of the observer’s inclination, which is different for each row: $\theta_0= \pi/2$ (top),  $\theta_0= \pi/6$ (center) and  $\theta_0= 0$ (bottom). As $\theta_0\to0$, the shadows become more circular and approach perfect centering, in agreement with the expected axial symmetry of the system. In particular,  Figs. \ref{shadow_throat}, \ref{shadow_throat_30}, and \ref{shadow_throat_0} exhibit no variation because for $a=0$ the configuration is spherically symmetric and all lines of sight are equivalent.

\section{Some representative examples}\label{sec:results}

In this section, we present the shadows obtained following the procedure explained in Sec. \ref{sec:shadow}, for particularly interesting values of the parameters. Furthermore, we compare the analytical shadows with those obtained by using the PyHole backward ray-tracing code \cite{pyhole}.

In the ray-tracing method, light rays are traced backward from the observer and are either scattered or absorbed by the black hole or wormhole. The ray paths are terminated when they reach an outer sphere---indicating that the rays have escaped---or when they cross the corresponding event horizon---indicating absorption. For visualization in the observer's sky, escaping rays are colored according to the quadrant of the outer sphere they reach, while absorbed rays are colored black, forming the shadow.\footnote{We also use a grid with 18 lines per $180^{\circ}$ in the observer sky for better visualization.} In our case, when no horizons are present, we consider a light ray to be absorbed---indicating that it crosses through the throat---if it reaches a cutoff radius, defined as $r_{\text{cutoff}}/m=1\times 10^{-3}$. For the numerical implementation, it was necessary to specify the conformal factor function $\Psi(r,\theta)$ in (\ref{rotating_lessa_olmo}). Following \cite{kocherlakota2023, kar2025}, we adopted the reasonable choice $\Psi=\rho^2$. For further details on the ray-tracing technique and the code used, we refer the reader to \cite{pyhole,other2015}.

\begin{figure}
    \centering
    \includegraphics[width=.9\textwidth]{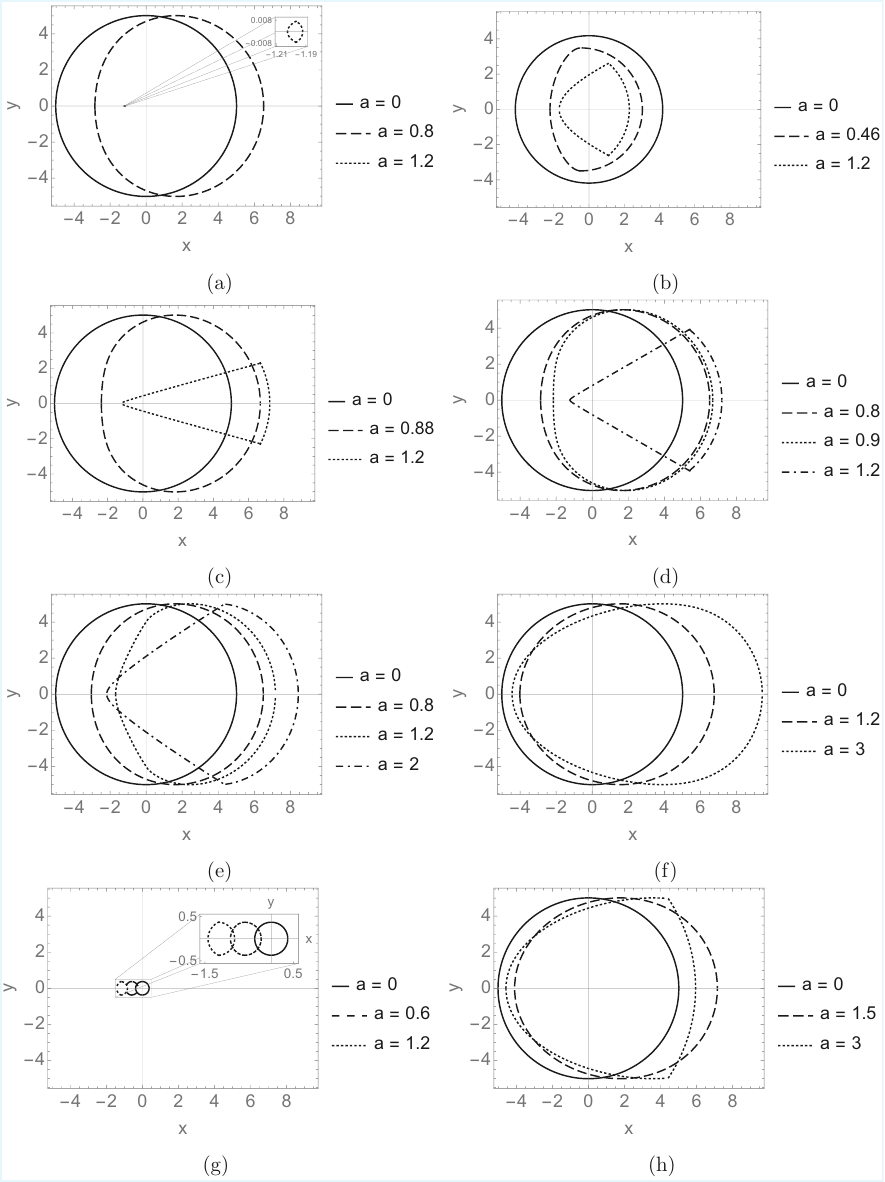}
    \begin{minipage}{0.45\textwidth}
      \phantomsubcaption\label{shadow_case1}
    \end{minipage}
    \begin{minipage}{0.45\textwidth}
      \phantomsubcaption\label{shadow_case1b}
    \end{minipage}
    \begin{minipage}{0.45\textwidth}
      \phantomsubcaption\label{shadow_case2}
    \end{minipage}
    \begin{minipage}{0.45\textwidth}
      \phantomsubcaption\label{shadow_case3}
    \end{minipage}
    \begin{minipage}{0.45\textwidth}
      \phantomsubcaption\label{shadow_case4}
    \end{minipage}
    \begin{minipage}{0.45\textwidth}
      \phantomsubcaption\label{shadow_case5}
    \end{minipage}
    \begin{minipage}{0.45\textwidth}
      \phantomsubcaption\label{shadow_case6}
    \end{minipage}
    \begin{minipage}{0.45\textwidth}
      \phantomsubcaption\label{shadow_case6b}
    \end{minipage}
    \caption{Contours of the shadows for the different cases with the observer on the equatorial plane. In all plots $\omega=m=1$. (\subref{shadow_case1}) Case 1 with $\rho_0=0.2$, $l=0.05$. (\subref{shadow_case1b}) Case 1 with $\rho_0=0.9$, $l=0.66$. (\subref{shadow_case2}) Case 2 with $\rho_0=0.2=\mathcal{F}(l\simeq0.106,\omega,m)$. (\subref{shadow_case3}) Case 3 with $\rho_0=0.2$, $l=0.15$. (\subref{shadow_case4}) Case 4 with $\rho_0=0.2$, $l=0.5$. (\subref{shadow_case5}) Case 5 with $\rho_0=0.2=\mathcal{F}(l\simeq1.894,\omega,m)$. (\subref{shadow_case6}) Case 6 with $\rho_0=1.2$, $l=0.5$. (\subref{shadow_case6b}) Case 6 with $\rho_0=0.2$, $l=2$. The value of $a$ for each curve is indicated at the right side of the plots.} 
    \label{fig:shadows}
\end{figure}

\begin{figure}
    \centering
    \includegraphics[width=.96\textwidth]{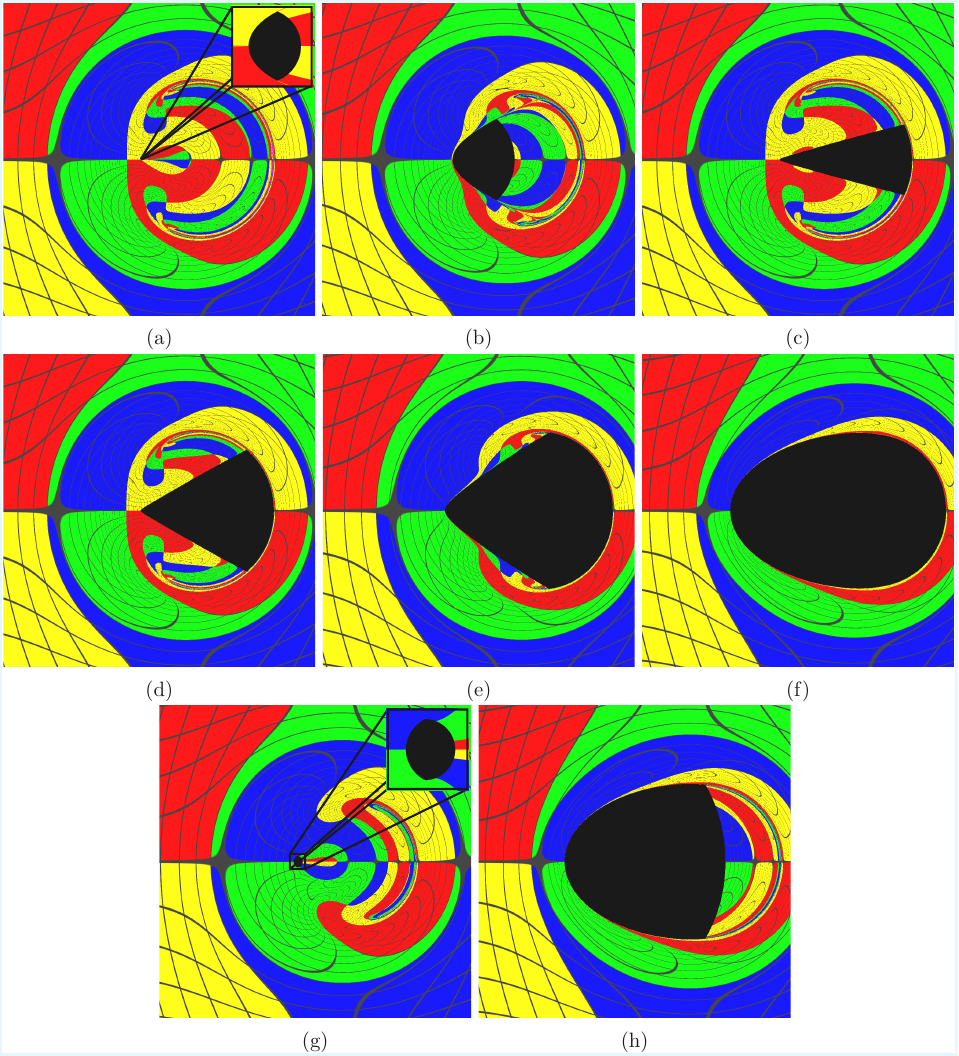}
    \begin{subfigure}{.32\linewidth}
        \phantomsubcaption\label{ray_tracing_case1}
    \end{subfigure}
    \begin{subfigure}{.32\linewidth}
        \phantomsubcaption\label{ray_tracing_case1b}
    \end{subfigure}
    \begin{subfigure}{.32\linewidth}
        \phantomsubcaption\label{ray_tracing_case2}
    \end{subfigure}
    \begin{subfigure}{.32\linewidth}
        \phantomsubcaption\label{ray_tracing_case3}
    \end{subfigure}
    \begin{subfigure}{.32\linewidth}
        \phantomsubcaption\label{ray_tracing_case4}
    \end{subfigure}
    \begin{subfigure}{.32\linewidth}
        \phantomsubcaption\label{ray_tracing_case5}
    \end{subfigure}
    \begin{subfigure}{.32\linewidth}
        \phantomsubcaption\label{ray_tracing_case6}
    \end{subfigure}
    \begin{subfigure}{.32\linewidth}
        \phantomsubcaption\label{ray_tracing_case6b}
    \end{subfigure}
    \caption{Shadows (black regions) for different cases with the observer on the equatorial plane, obtained numerically using the ray-tracing method. In all plots $\omega=m=1$. (\subref{ray_tracing_case1}) Case 1 with $\rho_0=0.2$, $l=0.05$, $a=1.2$. (\subref{ray_tracing_case1b}) Case 1 with $\rho_0=0.9$, $l=0.66$, $a=1.2$. (\subref{ray_tracing_case2}) Case 2 with $\rho_0=0.2=\mathcal{F}(l\simeq0.106,\omega,m)$, $a=1.2$. (\subref{ray_tracing_case3}) Case 3 with $\rho_0=0.2$, $l=0.15$, $a=1.2$. (\subref{ray_tracing_case4}) Case 4 with $\rho_0=0.2$, $l=0.5$, $a=2$. (\subref{ray_tracing_case5}) Case 5 with $\rho_0=0.2=\mathcal{F}(l\simeq1.894,\omega,m)$, $a=3$. (\subref{ray_tracing_case6}) Case 6 with $\rho_0=1.2$, $l=0.5$, $a=1.2$. (\subref{ray_tracing_case6b}) Case 6 with $\rho_0=0.2$, $l=2$, $a=3$.}
    \label{fig:ray_tracing}
\end{figure}

Given the behavior shown in Fig. \ref{fig:shadows_examples}, we now focus on the case of an observer located in the equatorial plane, where for a certain value of the rotation parameter $a\neq 0$ the deformation of the shadow is most prominent. We present our results graphically: Fig. \ref{fig:shadows} displays the contours of the shadows obtained for different values of the parameter $a$ in each of the cases detailed in Sec. \ref{sec:rotating_black_bounces}, while Fig. \ref{fig:ray_tracing} shows the corresponding ray-tracing image for a representative value of $a$ in each case. We can see that:
\renewcommand{\theenumi}{\roman{enumi}}
\begin{enumerate}
    \item For case 1 (Fig. \ref{shadow_case1}), the shadow has an abrupt change in size when the black bounce loses its horizon, because at that point the shadow goes from being formed solely by the photon sphere present outside the throat (which from here forms an opened curve, like in the Kerr case \cite{kerr_naked}) to being closed by the wormhole throat, which now forms the rest of the shadow boundary. As a consequence, its size is reduced to 3 orders of magnitude. The resulting image obtained with the ray-tracing method for this case with a very small shadow is shown in Fig. \ref{ray_tracing_case1}.
    \item For cases 2, 3, and 4 (Figs. \ref{shadow_case2}, \ref{shadow_case3}, and \ref{shadow_case4}, respectively), the shadow becomes increasingly deformed and shifted to the right as $a$ increases, as in the Kerr case, but for large $a$, it takes on a cusped, trianglelike shape with a rounded side. These cusped shadows are not new for rotating wormholes, as can be seen in \cite{throat_1,throat_4,cusped_3,cusped_4}. Before the cusped form, there is a D-shaped shadow for a critical value of $a$ in cases 2 and 3, but not in case 4. In all these cusped cases, the curves consist of two distinct parts arising from different and disconnected ranges of the coordinate $r$ of closest approach, corresponding to two separate families of unstable spherical orbits. Both curves are large, but we cut them at their intersection, which represents the shadow boundary, as confirmed by the ray-tracing method in Figs. \ref{ray_tracing_case2}, \ref{ray_tracing_case3}, and \ref{ray_tracing_case4}. In case 4, the mechanism of formation of the cusp form is the same as described in \cite{throat_1} and \cite{throat_4} for other rotating wormholes, with a family of stable spherical orbits connecting the two unstable ones. In cases 2 and 3, there is no patch between the curves.
    \item For case 5 (Figs. \ref{shadow_case5} and \ref{ray_tracing_case5}), we always have closed and smooth shadows that become elongated for large $a$. This is similar to what happens with the topologically charged rotating wormhole described in \cite{elongated}.
    \item For case 6 with $\rho_0>\mathcal{F}_{max}$ and $\omega>1/2$ (Figs. \ref{shadow_case6} and \ref{ray_tracing_case6}), we obtain small shadows (similar to those found in case 1 at large $a$), which always include a contribution from the wormhole throat, regardless of the value of $a$. This is because the contribution of the photon sphere present outside the throat either forms open curves or, in the case of $a=0$, is entirely absent (as can be seen from Eq. (\ref{static_shadow_example}), which has no solutions for $\rho_0>9/8$). These unusually small wormhole shadows resemble those described in Ref. \cite{wang2020} for an asymmetric thin-shell wormhole, which could be arbitrarily small.
    \item Finally, for case 6 with $[(\mathcal{F}(l,\omega>1/2,m)<\rho_0\leq\mathcal{F}_{max})\text{ and }\,l>l_{max}]$ or $\rho_0<\mathcal{F}(l,0<\omega<1/2,m)$ (Figs. \ref{shadow_case6b} and \ref{ray_tracing_case6b}), we have something very similar to case 5, with shadows that become elongated with increasing $a$, until a critical value of $a$ when the two curves corresponding to the possible contributors (the throat and the photon sphere present outside the throat) intersect. So, from here and for large values of $a$, the shadows have two contributions, as we have seen for the case described in Fig. \ref{shadow_both}. This also happens for some values of the parameters for case 1, when we are near case 2 in the corresponding ``phase diagram," as can be seen in Figs. \ref{shadow_case1b} and \ref{ray_tracing_case1b}.
\end{enumerate}

\section{Final remarks}\label{sec:final_remarks}

In this article, we have analyzed the shadows produced by the rotating version of a black bounce spacetime proposed by Lessa and Olmo \cite{lessaolmo2025}. First, we have obtained the metric and we have distinguished between scenarios with and without event horizons, as they require separate analyses. In cases with at least one horizon, the shadow boundary is formed by light rays that originate from the photon sphere. In contrast, for horizonless configurations, the boundary is shaped either by rays that would originate from a photon sphere present outside the throat---if such a structure exists---or by rays that have a turning point at the wormhole throat, or by a combination of both.

It is important to note that, in contrast to known rotating wormholes, the contribution from the throat to the shadow boundary in the horizonless cases of our metric is not due to the presence of a photon sphere located at the throat. Instead, as the impact parameters change, the turning point of the light-ray trajectories approaches the throat, due to the presence of a potential barrier located there. At a critical value, the ray crosses the throat without returning, contributing to the shadow interior. This behavior is not associated with a maximum of the effective potential at the throat---as would be expected for a photon sphere---but rather with a discontinuity, where the potential abruptly drops and exhibits only a root there.

Our analytical study is valid for any value of the inclination angle of the observer. Considering a fixed value of $a\neq 0$, when $\theta_0= 0$ (polar observer) the shadow is circular while for $\theta_0 \neq 0$ it has a deformation that grows with $\theta_0$, being the effect most prominent when $\theta_0= \pi /2$ (equatorial observer). Then, we have concentrated in the equatorial scenario in our examples. Based on the values of the metric parameters, we have identified five distinct types of shadows.
\begin{enumerate}
    \item First, we find cases where, when the values of the rotation parameter $a$ are not very large, the shadows closely resemble those of the Kerr black hole. For large enough $a$---where the Kerr shadow would disappear---a small shadow remains, closed by the contribution from the wormhole throat (Fig. \ref{shadow_case1}).
    \item In some configurations, for large enough $a$, the shadow takes a cusped, triangularlike shape (Figs. \ref{shadow_case2}, \ref{shadow_case3}, and \ref{shadow_case4}).
    \item Other cases exhibit elongated shadows at high enough rotation (Fig. \ref{shadow_case5}).
    \item In certain scenarios, the shadow remains small across all values of $a$, always with a contribution from the wormhole throat (Fig. \ref{shadow_case6}).
    \item Finally, for large enough $a$, some cases feature shadows formed by both contributions---from the throat and from the photon sphere present outside the throat---resulting in a mixed boundary structure, but without the sharp reduction in size observed in the previous cases (Figs. \ref{shadow_case1b} and \ref{shadow_case6b}).
\end{enumerate}

Among the different types of shadows analyzed, those with an elongated or triangularlike shape (in addition to the Kerr-like ones that appear when the values of the rotation parameter $a$ are not very large) have known precedents in the literature for other black hole, wormhole, or black bounce metrics. However, the shadows obtained in cases 1 and 6, where the throat contributes to their formation, exhibit genuinely novel features. Their silhouettes may resemble previously known profiles but appear abruptly truncated due to the throat contribution, or they may result in remarkably small shadows for the parameters considered. This behavior can be understood from the fact that the mechanism by which the throat contributes to these shadows is fundamentally different from those previously discussed, as it does not arise from the presence of a photon sphere. In this sense, the results presented here constitute a new development in this line of research, particularly regarding the role that a wormhole throat can play in shaping the shadow.

\section*{Acknowledgments}

This work has been supported by CONICET.

\end{document}